\documentclass[onecolumn,A4]{article}

\pdfoutput=1
\usepackage{graphicx}
\usepackage{epsfig}
\usepackage{subfigure}
\usepackage{cite}
\usepackage{anysize}
\usepackage{amssymb}
\usepackage{amsthm}

\begin{document}

\vskip 1cm

\begin{center}

{\bf{\large Comparison of Bulk Micromegas with Different Amplification Gaps}} \\
~\\
Purba Bhattacharya$^{*,1}$, Sudeb Bhattacharya$^{1}$, Nayana Majumdar$^{1}$, Supratik Mukhopadhyay$^{1}$, Sandip Sarkar$^{1}$, Paul Colas$^{2}$, David Attie$^{2}$
~\\
{\em $^{1}$Applied Nuclear Physics Division, Saha Institute of Nuclear Physics, Kolkata - 700064, India  \\
$^{2}$DSM/IRFU, CEA/Saclay, F-91191 Gif-sur-Yvette CEDEX, France}
~\\
~\\
~\\
~\\
~\\
{\bf{\large Abstract}}
\end{center}

The bulk Micromegas detector is considered to be a promising candidate for building TPCs for several future experiments including the projected linear collider.
The standard bulk with a spacing of $128~\mu\mathrm{m}$ has already established itself as a good choice for its performances in terms of gas gain uniformity, energy and space point resolution, and its capability to efficiently pave large readout surfaces with minimum dead zone.  
The present work involves the comparison of this standard bulk with a relatively less used bulk Micromegas detector having a larger amplification gap of $192~\mu\mathrm{m}$. 
Detector gain, energy resolution and electron transparency of these Micromegas have been measured under different conditions in various argon based gas mixtures to evaluate their performance. 
These measured characteristics have also been compared in detail to numerical simulations using the Garfield framework that combines packages such as neBEM, Magboltz and Heed.
Further, we have carried out another numerical study to determine the effect of dielectric spacers on different detector features.
A comprehensive comparison of the two detectors has been presented and analyzed in this work.

\vskip 1.5 cm
\begin{flushleft}
{\bf Keywords}: Bulk Micromegas, Electron Transparency, Gain, Energy Resolution, Dielectric Spacer

\end{flushleft}

\vskip 3in
\noindent
{\bf ~$^*$Corresponding Author}: Purba Bhattacharya

Electronic mail: purba.bhattacharya@saha.ac.in

\section{Introduction}
\label{sec: introduction}

A large high performance TPC has been proposed for the study of physics in the future linear collider.
In order to achieve better resolution with the capability of handling high luminosity, Micro-Pattern Gaseous Detectors (MPGDs) have been proposed to be used instead of Multi Wire Proportional Counter (MWPC). 
These next generation TPC detectors will have full readout  coverage to provide excellent overall pattern recognition and tracking.
The Micromegas (MICRO-MEsh GAseous Structure), under development for the future TPC \cite{TPC}, is a parallel plate device and composed of a very thin metallic micro-mesh, which separates the low field drift region from the high-field amplification region.
Amplification gaps of $50 - 150~\mu\mathrm{m}$ allow fast evacuation and is found to
work optimally for most experiments.
The bulk Micromegas detector \cite{Bulk} with a standard amplification gap of $128~\mu\mathrm{m}$ is already known to be a good candidate for a readout system in different TPCs for its performance in terms of gas gain uniformity, energy resolution and its capability to efficiently pave large readout surfaces with minimum dead zone.
For experiments involving low pressure operation, Micromegas detectors having
larger amplification gap are more useful because they allow sufficient gain
despite longer ionization mean free path of electrons.
Several rare event detection experiments are known to be using
Micromegas with amplification gaps of around $200~\mu\mathrm{m}$.

In this work, we will discuss the experimental and numerical studies carried out for the characterization of less used bulk Micromegas having larger amplification gap of $192~\mu\mathrm{m}$.
A comprehensive comparison of several detector characteristics of this large gap Micromegas with a standard $128~\mu\mathrm{m}$ bulk in several argon based gas mixture has been carried out.
It should be mentioned that the same experimental set up and numerical simulation framework have been seamlessly used for both the Micromegas detectors. 
Finally a brief numerical study on the effect of dielectric spacers on electric field, gain and detector signal has been discussed.

\section{Experimental set-up}
\label{sec: experiment}

\begin{figure}[hbt]
\centering
\subfigure[]
{\label{Microscope}\includegraphics[scale=0.25]{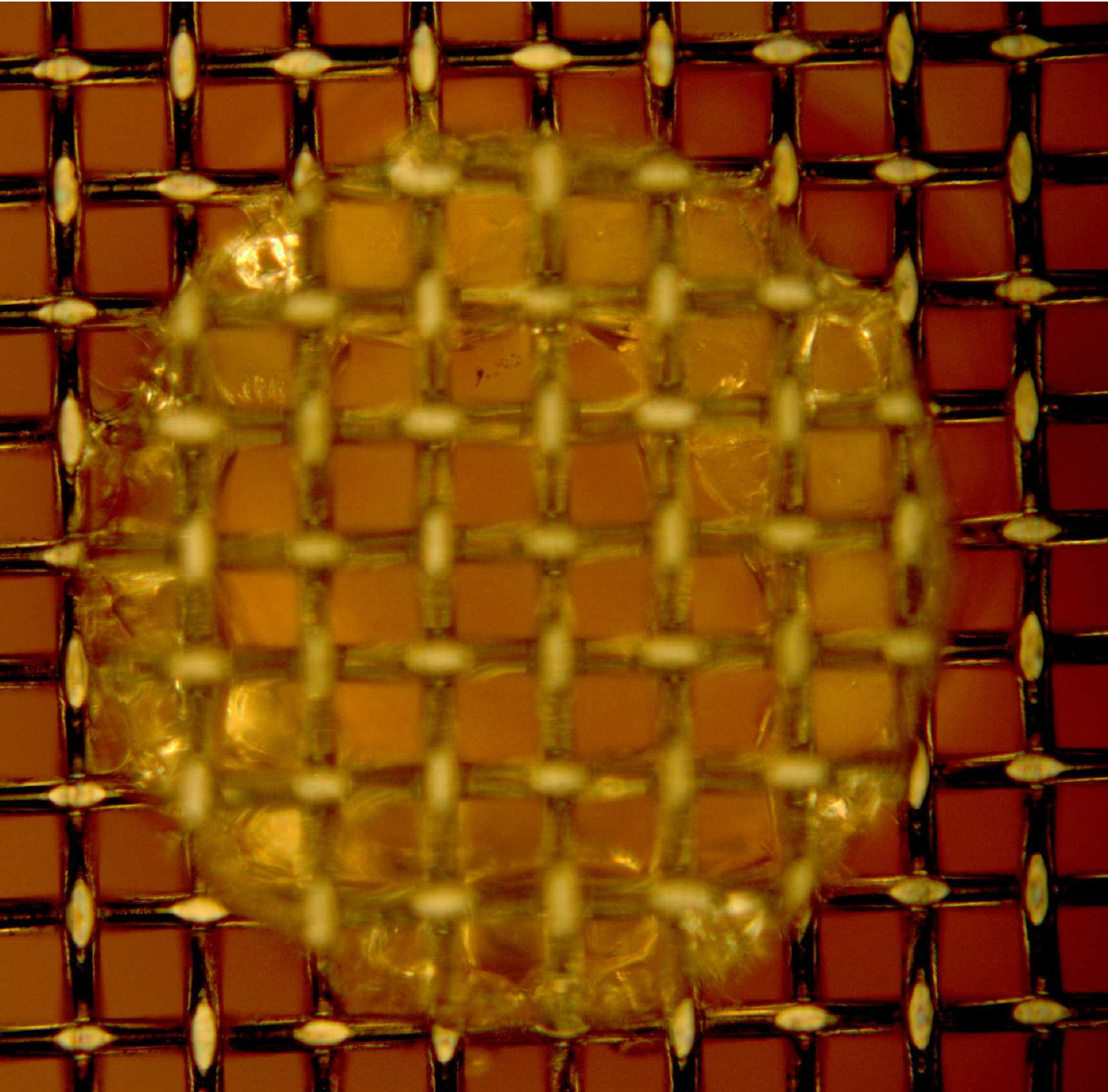}}
\subfigure[]
{\label{Simulation}\includegraphics[scale=0.525]{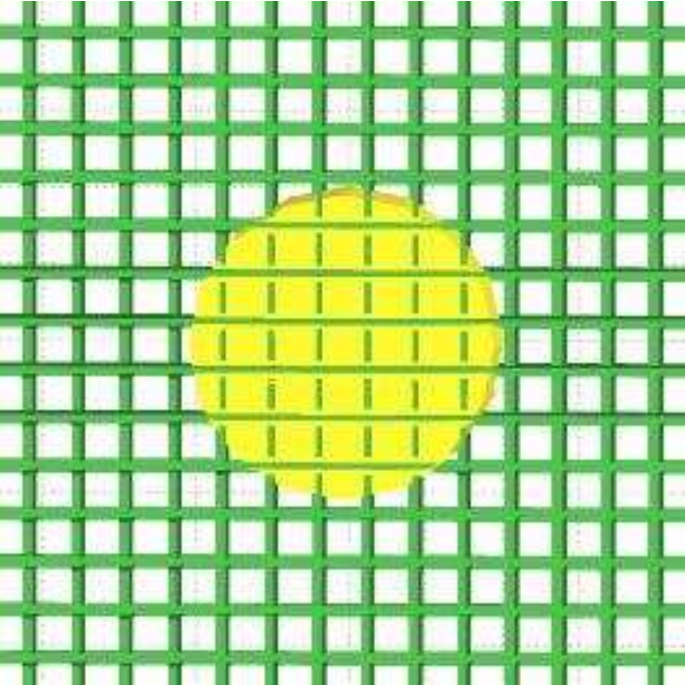}}
\caption{Micro-mesh: (a) Microscopic view, (b) Simulated model}
\label{Micromesh}
\end{figure}

The prototype detectors with an active area of $15~\mathrm{{cm}^2}$ have been fabricated in Saclay, France and tested in Kolkata, India.
The bulk Micromegas detectors were equipped with a calendered woven micro-mesh similar to that of the T2K experiment (Fig.~\ref{Micromesh}). 
The stainless-steel wires had a diameter of $18~\mu\mathrm{m}$ with a pitch of $63~\mu\mathrm{m}$.
The anode of both the detectors were non-segmented.
The drift distance was $1.2~\mathrm{cm}$.
The chamber was flushed with different argon based gas mixtures at room temperature (296 K) and 1 atmospheric pressure.
The detectors were tested by means of X-ray quanta from a $^{55}\mathrm{Fe}$ source.
The output was passed through a charge sensitive pre-amplifier (ORTEC model 142IH).
Subsequently, it was fed to a spectroscopic amplifier (ORTEC model 672) with an integration time of $1~\mu\mathrm{sec}$. 
Finally, the data were recorded in a AMTEK MCA 800A.

\section{Simulation Tools}
\label{sec: simulation}

The experimental data were compared with estimates obtained through numerical simulation. 
We have used the Garfield \cite{Garfield} simulation framework.
This framework was augmented in 2009 through the addition of the neBEM \cite{neBEM} toolkit to carry out 3D electrostatic field simulation.
Earlier, Garfield had to import field maps from one of the several commercial FEM packages in order to study 3D gas detectors.
Due to the exact foundation expressions based on the Green's functions, the neBEM approach has been found to be exceptionally accurate in the complete physical domain, including the near field.
This fact, in addition to other generic advantages of BEM over FEM, makes neBEM a strong candidate as a field-solver for MPGD related computations.
Besides neBEM, the Garfield framework provides interfaces to HEED \cite{Heed} for primary ionization calculation and Magboltz \cite{Magboltz} for computing drift, diffusion, Townsend and attachment coefficients. 

\section{Results}
\label{sec: results}

\subsection{Electron Transparency of Micro-mesh}
\label{sec: transparency}

\begin{figure}[hbt]
\centering
\subfigure[]
{\label{Trans-compare}\includegraphics[scale=0.35]{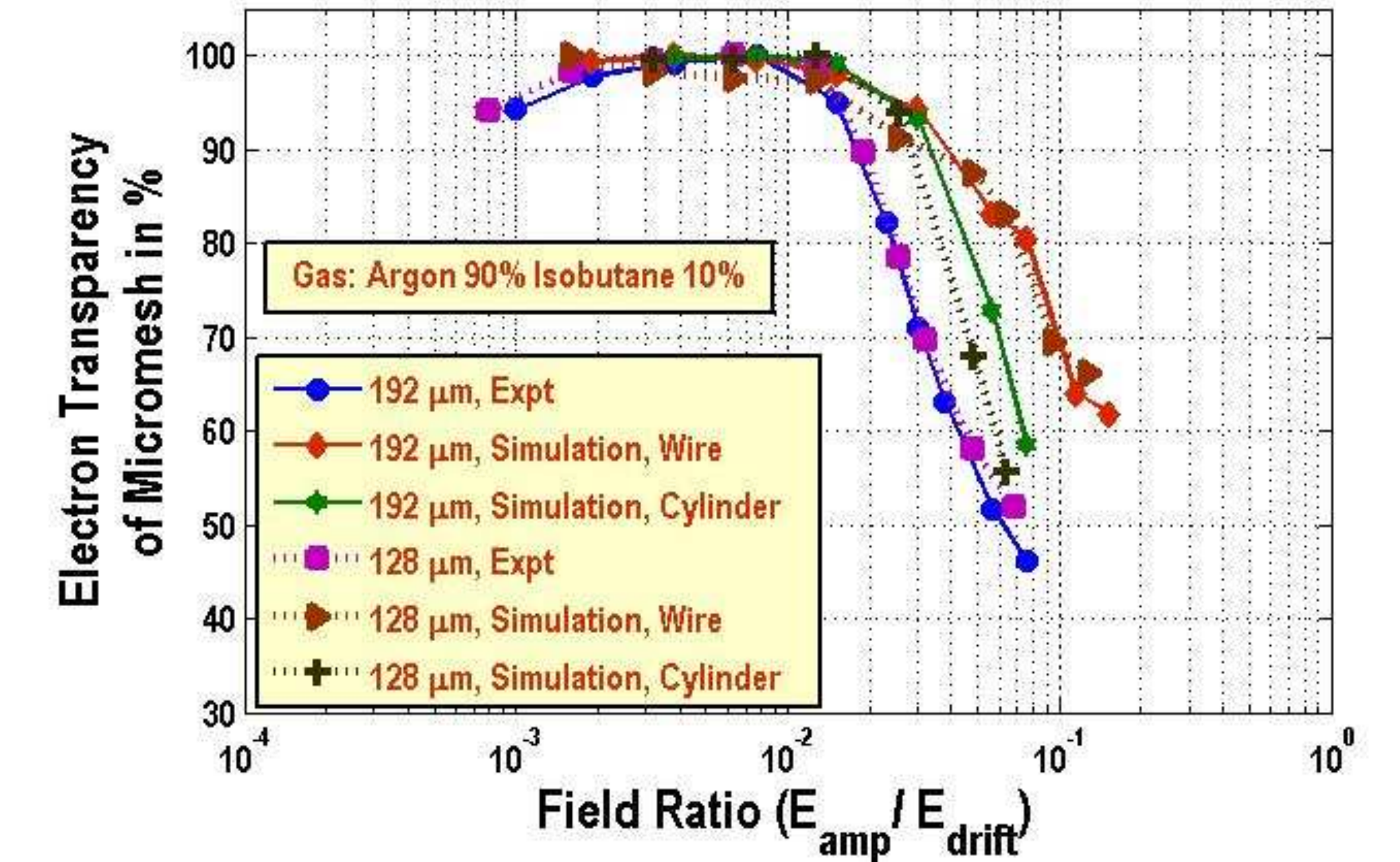}}
\subfigure[]
{\label{Contour}\includegraphics[scale=0.35]{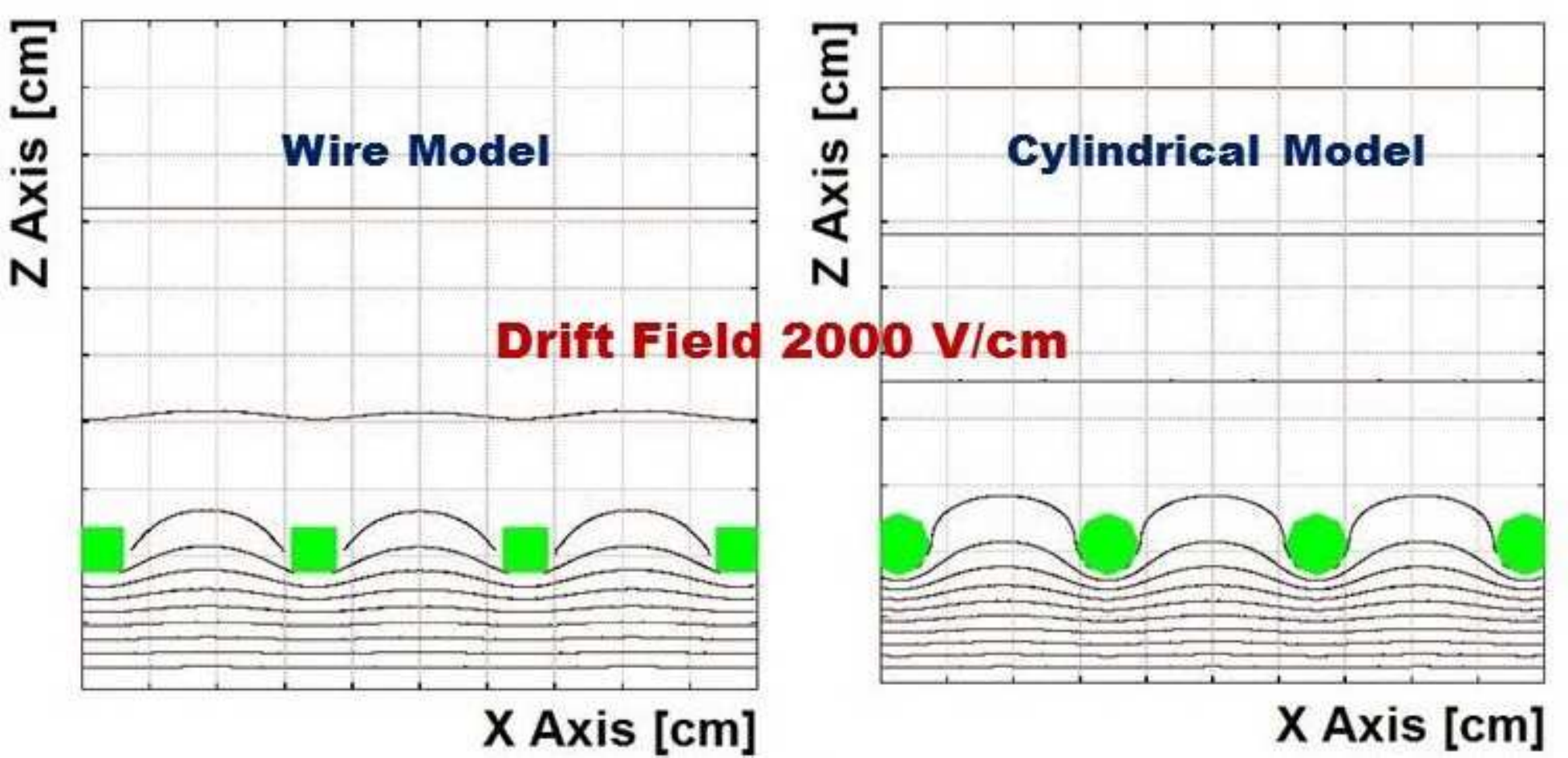}}
\caption{(a) Comparison of measured and simulated transparency obtained for both cylindrical and wire model, (b) Equipotentials in the mid-plane of a mesh, thin wire approximation and octagonal approximation of solid cylinder}
\label{Transparency}
\end{figure}

We estimate the mesh transparency as the ratio of the signal amplitude at a given drift field over the signal amplitude at drift field $100~\mathrm{V/cm}$ and $200~\mathrm{V/cm}$ for $192~\mu\mathrm{m}$ and $128~\mu\mathrm{m}$, respectively.
For both the cases, the measured electron transparency is almost constant for field ratio ($E_\mathrm{drift}/E_\mathrm{amp}$) $< 0.01$, beyond which electrons increasingly hit the mesh and decreases to $50~\%$ at field ratio 0.006.
The measured values and simulated estimates of electron transparency are compared in Fig.~\ref{Trans-compare} for two different amplification gaps. 
These measurements are sensitive to electron transport at the micron-scale, and thus we use the microscopic tracking method for the numerical process.
The transparency was calculated by drifting the electrons from randomly distributed points in the drift region, $100~\mu\mathrm{m}$ from the mesh and was estimated as the fraction of electrons arriving in the amplification region.
We use two different models with which a mesh can be modelled: one-dimensional thin wire segments and three-dimensional polygonal approximations of cylinders.
The wire model is computationally attractive but the field is calculated using the thin-wire approximation.
As a result, the voltage at one wire radius from the axis will only on an average equal to the imposed surface potential.
When cylindrical model is used, the voltage boundary condition is applied to each surface panels of the cylinder.
Also, the thin-wire approximation neglects the dipole moment, created to ensure an equal potential on both the surfaces.
As illustrated in Fig.~\ref{Contour}, the calculation using thin-wire approximation miscalculates the potential in higher drift field.
As a result, at higher drift field, the cylindrical model allows better estimates of transparency.
The slight discrepancy between the cylindrical model and the experiment is possible due to less statistics.

\subsection{Gain}
\label{sec: gain}

\begin{figure}[hbt]
\centering
{\includegraphics[scale=0.35]{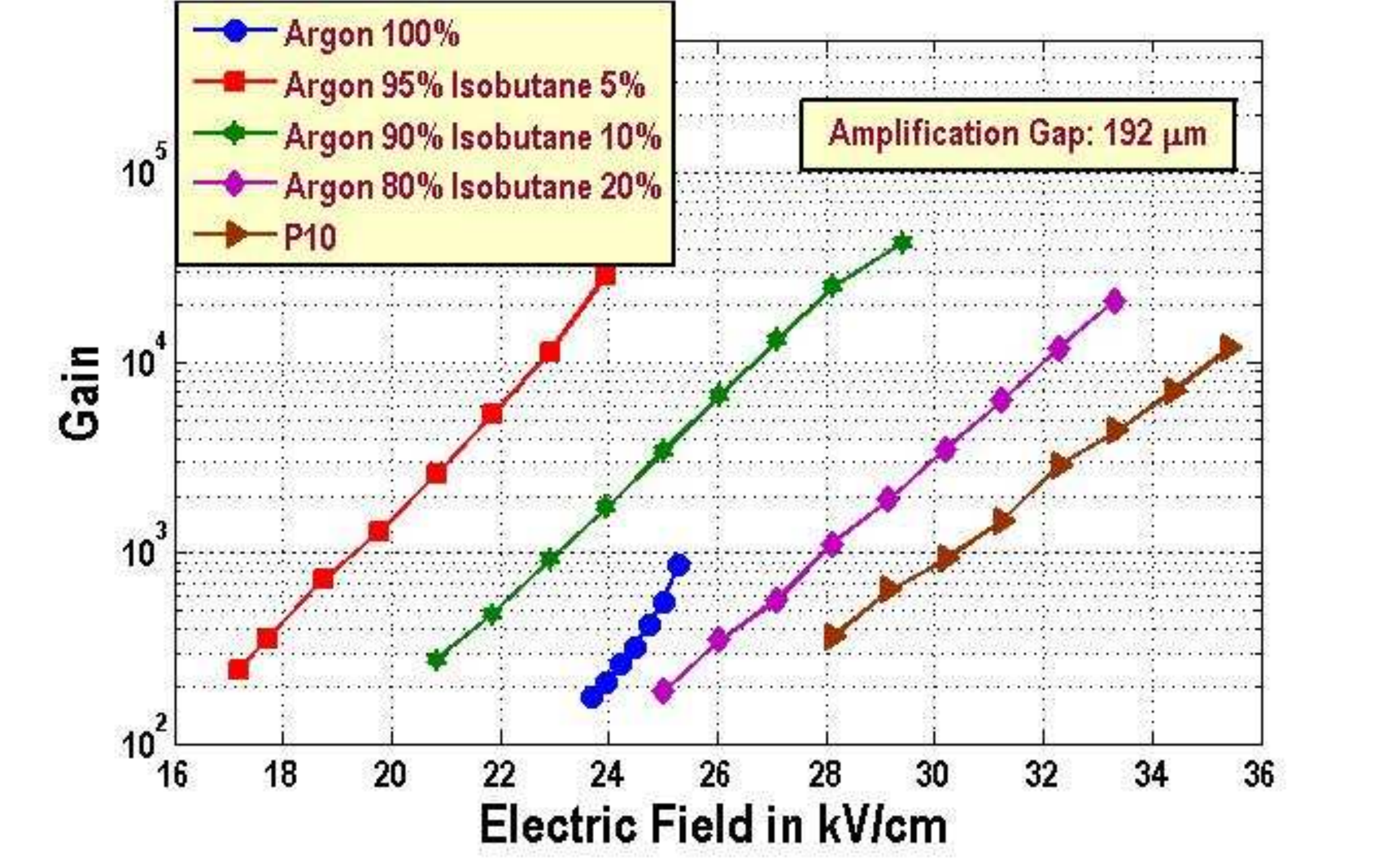}}
\caption{Measured gain for $192~\mu\mathrm{m}$}
\label{Gain-DiffGas}
\end{figure}

\begin{figure}[hbt]
\centering
{\includegraphics[scale=0.35]{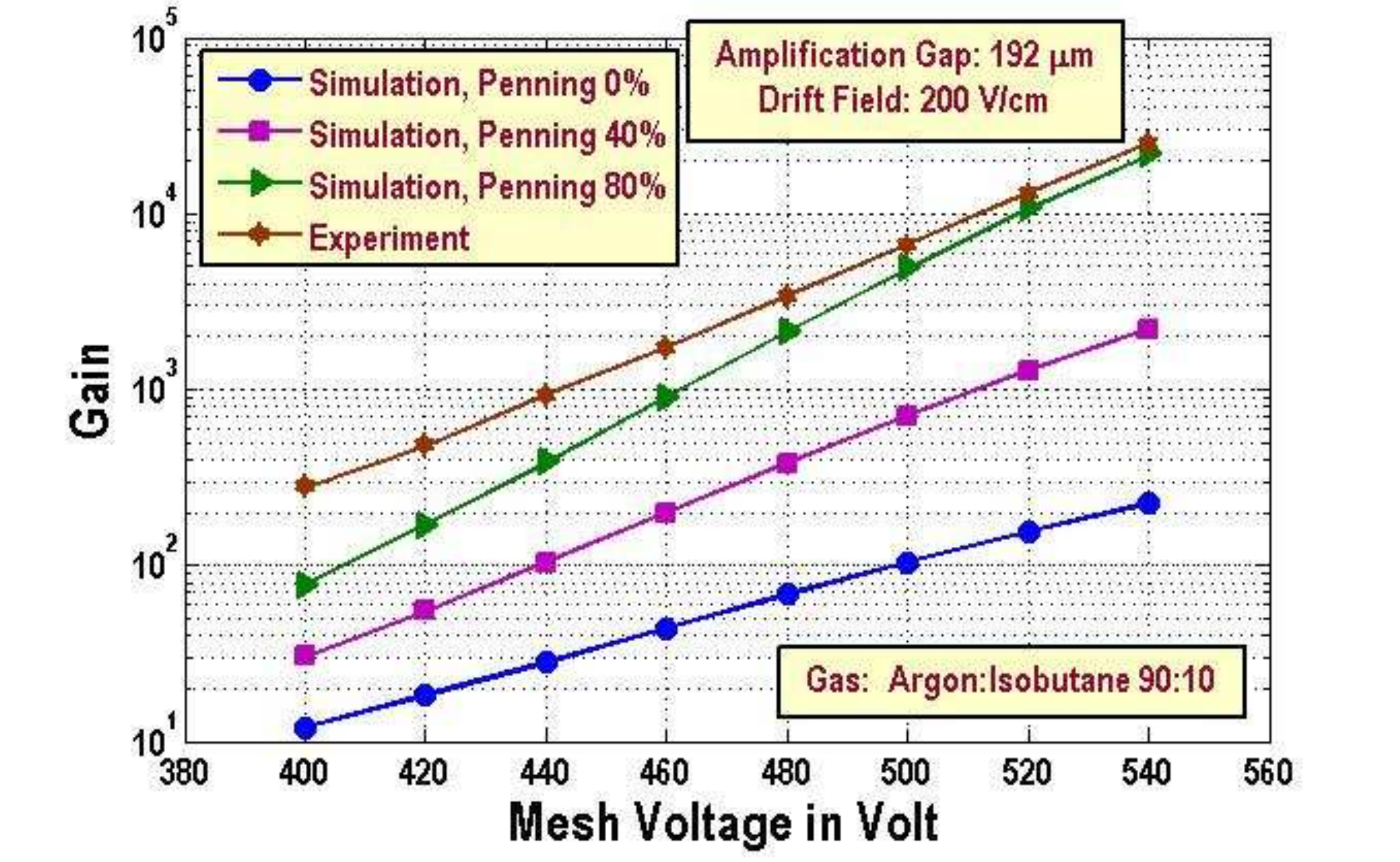}}
\caption{Comparison for Argon-Isobutane (90:10)}
\label{Gain-ArIso-Penning}
\end{figure}

\begin{figure}[hbt]
\centering
{\includegraphics[scale=0.35]{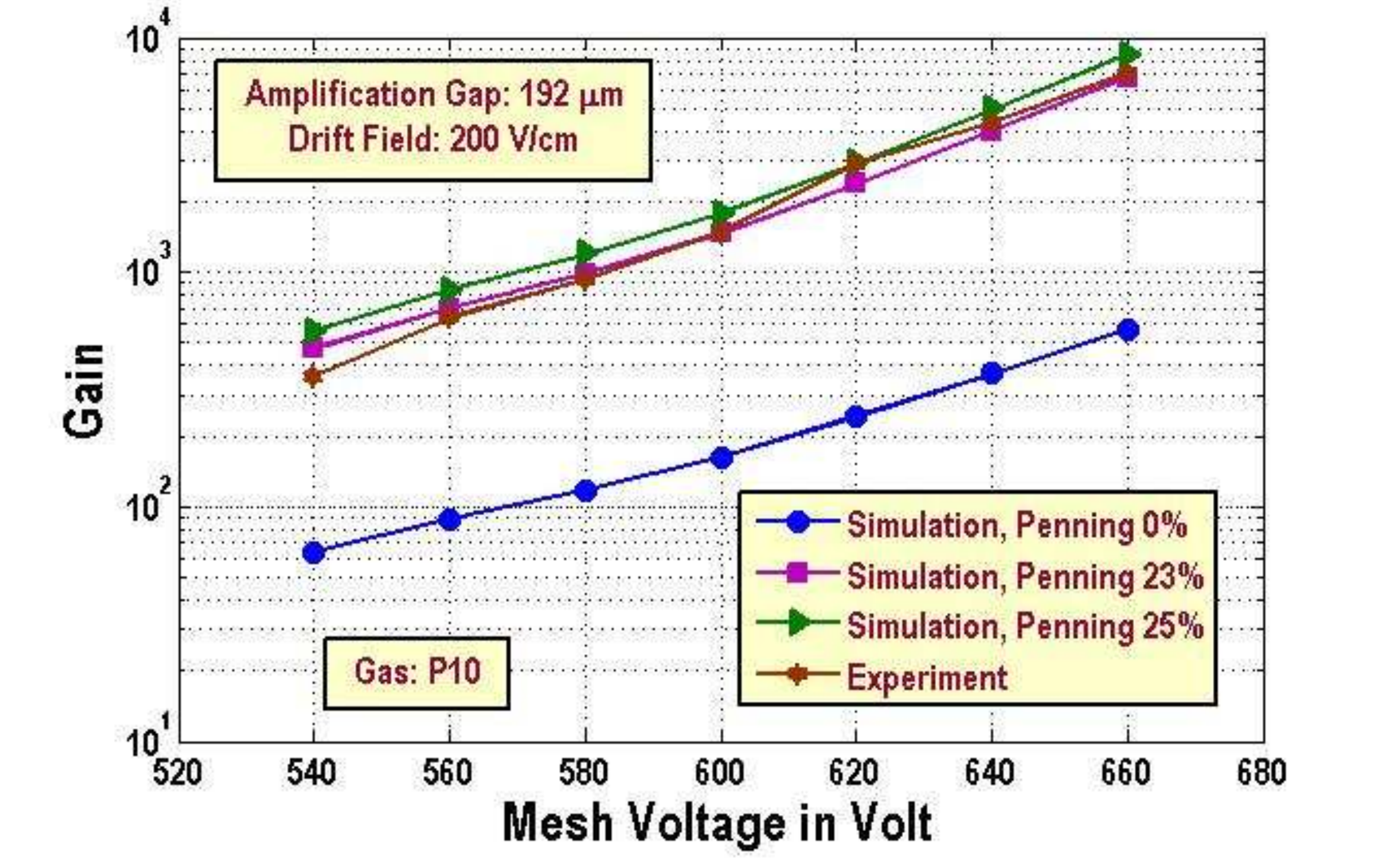}}
\caption{Comparison for P10}
\label{Gain-P10-Penning}
\end{figure}

The experimental gain curve for $192~\mu\mathrm{m}$ bulk for different argon-based gas mixtures is presented in Fig.~\ref{Gain-DiffGas}.
The plot shows that the maximum gain for Argon-Isobutane mixture is $3 \times 10^4$, whereas the maximum gain in P10, reaches to $1 \times 10^4$, at much higher electric field (36 kV/cm). 
It may be mentioned here that there was significant increase in the count rate for the $192~\mu\mathrm{m}$ bulk.
For numerical simulation, the effective gain of electrons is obtained as 
\begin{eqnarray}
\mathrm{g}_\mathrm{eff} = \eta \times \mathrm{g}_\mathrm{mult} \end{eqnarray}

\noindent where $\eta$ is the probability for a primary
electron to reach the amplification region i.e the electron transparency.
$\mathrm{g}_\mathrm{mult}$ is the multiplication factor of the
electrons throughout their trajectories.
The measurements and simulations are compared in Fig.~\ref{Gain-ArIso-Penning} and Fig.~\ref{Gain-P10-Penning}. 
The argon-based gas mixtures are Penning mixture. 
After considering results using different transfer rates, we chose to carry out the calculations with $80\%$ transfer rate for Argon-Isobutane mixture (90:10) and $25\%$ for P10 since these values agreed well with the experimental data.
Though the chosen transfer rate for P10 agreed well with that given in ref. \cite{Penning}, the transfer rate for Argon-Isobutane (90:10) was much higher than that predicted in \cite{Penning}.

\begin{figure}[hbt]
\centering
{\includegraphics[scale=0.35]{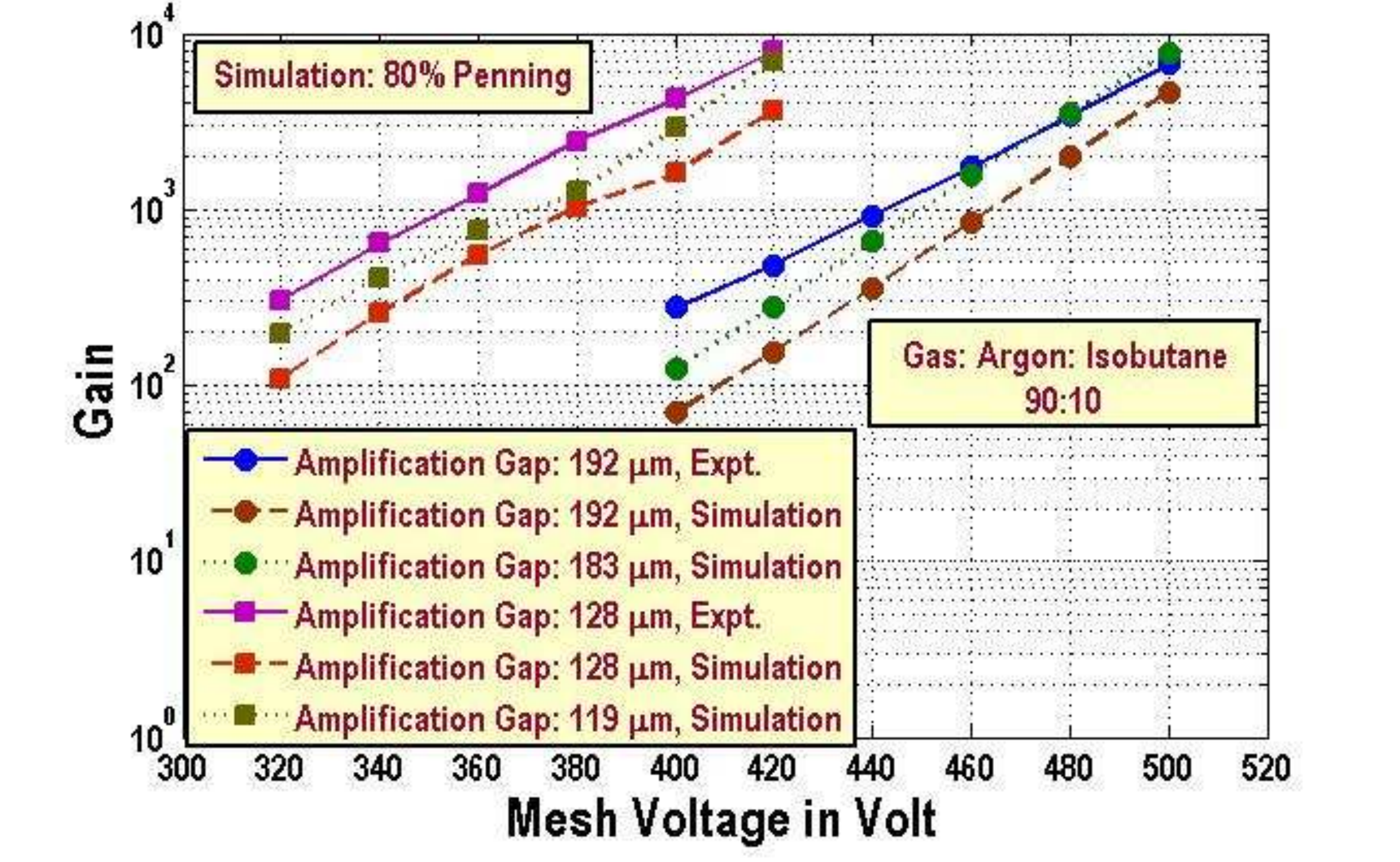}}
\caption{Comparison with $128~\mu\mathrm{m}$ bulk, (experiment and simulation)}
\label{Gain-DiffGap}
\end{figure}

A comparison with the $128~\mu\mathrm{m}$ detector in Argon-Isobutane (90:10) reveals that maximum experimental gain achieved with a larger gap is similar or slightly more than that with a smaller gap (Fig.~\ref{Gain-DiffGap}).
Though the simulation results (with Penning transfer rate $80\%$) for both detectors follow the experimental trend, the calculated values are still lower than the measured data.
One of the possible reasons may be that due to the fabrication process, the gap between the mesh plane and the anode plane is not exactly $128~\mu\mathrm{m}$ or $192~\mu\mathrm{m}$, rather less.
A small change in the amplification gap can affect the experimental gain considerably.
We have done a brief numerical study to investigate this effect.
We change the gap to $183~\mu\mathrm{m}$ and $119~\mu\mathrm{m}$ instead of $192~\mu\mathrm{m}$ and $128~\mu\mathrm{m}$ respectively, and recalculate the gain.
The gain increases considerably as seen in Fig.~\ref{Gain-DiffGap}.
This effect is more prominent in the case of smaller gap.

\subsection{Energy Resolution}
\label{sec: resolution}

\begin{figure}[hbt]
\centering
{\includegraphics[scale=0.35]{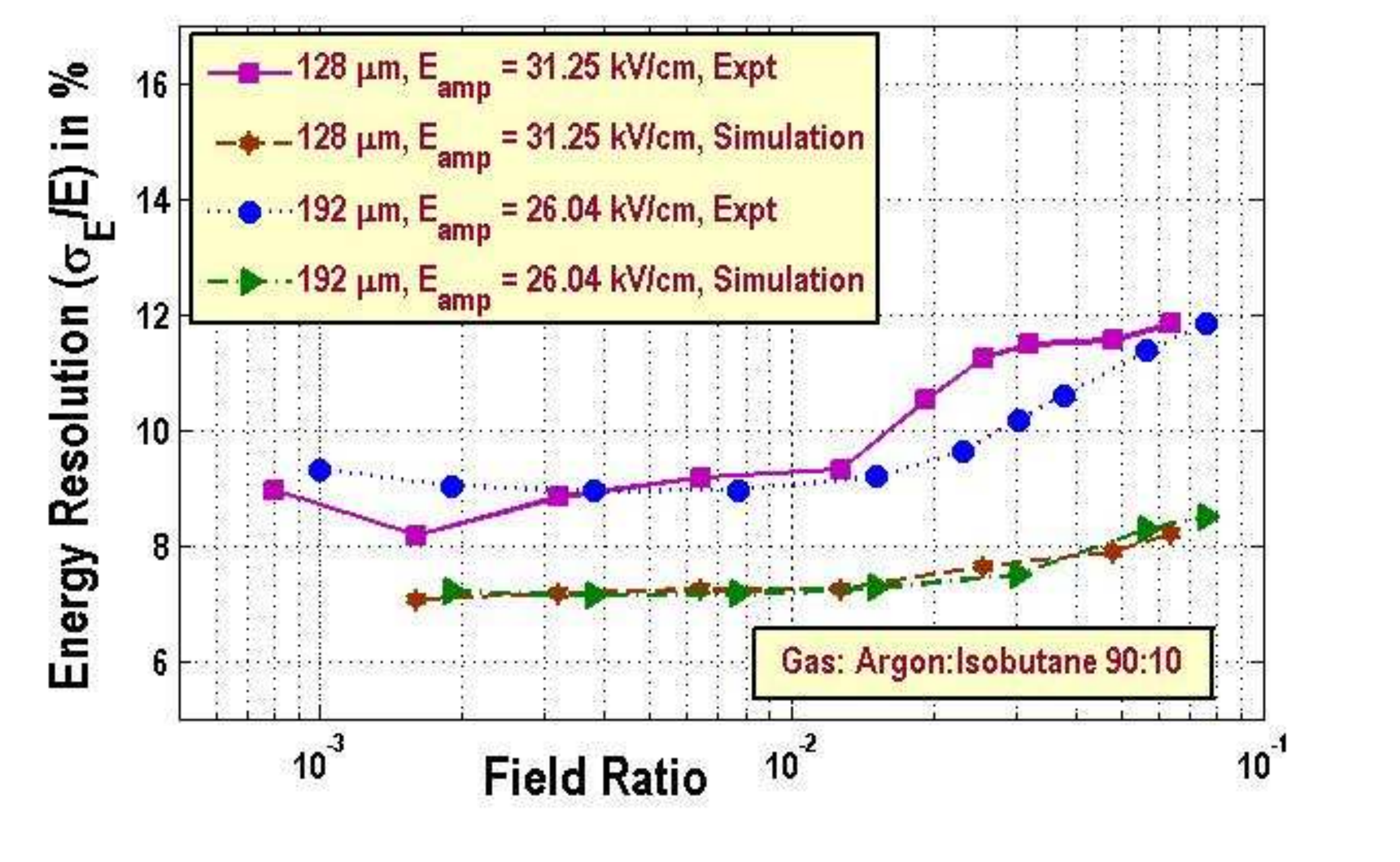}}
\caption{Variation of energy resolution with drift field}
\label{Energy Resolution}
\end{figure}

From Fig.~\ref{Energy Resolution} it can be seen that, at higher field ratios, the $192~\mu\mathrm{m}$ detector shows better resolution than the standard $128~\mu\mathrm{m}$ one.
For the numerical estimation of the energy resolution, we have followed the equation
\begin{eqnarray}
\mathrm{R^2} = (1/\mathrm{N}) \times (\mathrm{F} - 1 +(\mathrm{b+1})/\eta) 
\end{eqnarray}

\noindent where $\mathrm{R}$ represents the resolution, $\mathrm{N}$ is the primary electrons, $\mathrm{F}$ is the Fano Factor, and $\mathrm{b}$ is the relative variance of the gain distribution.
The calculated trend follows the measured data.
One of the reasons of the discrepancy between the experimental and simulated data can be the over estimation of the transparency using wire model as discussed in section \ref{sec: transparency}.
The gain variation also needs further investigation.

\section{Effect of dielectric spacer}
\label{sec: spacer}

In the practical realization of the Micromegas detectors, a set of dielectric spacers is required to keep a constant gap between the mesh and the anode.
We have studied the effect of such dielectric spacers (dielectric constant = 4) on different detector characteristics for a standard $128~\mu\mathrm{m}$ bulk.

\begin{figure}[hbt]
\centering
\subfigure[]
{\label{Field1}\includegraphics[scale=0.15]{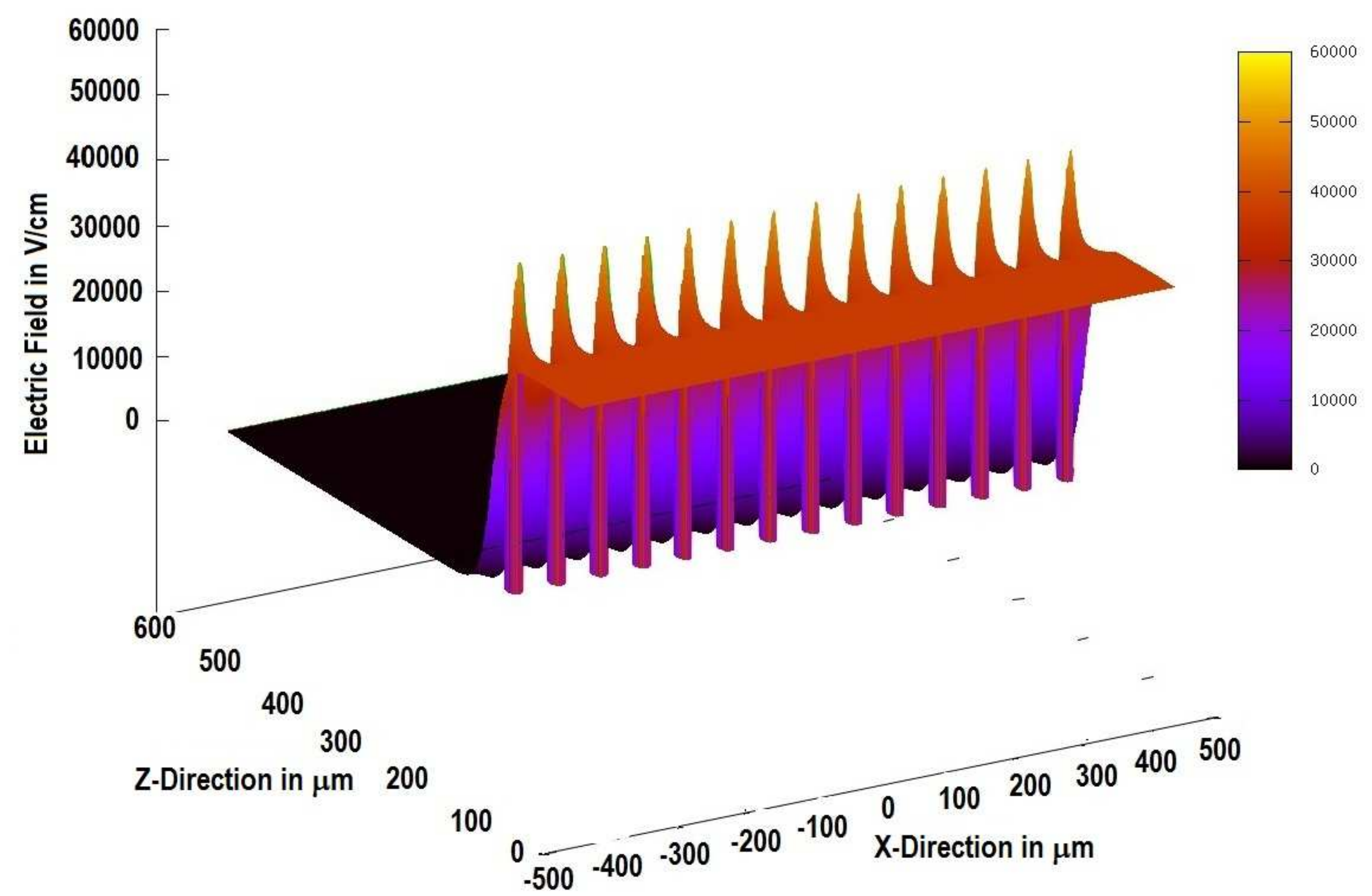}}
\subfigure[]
{\label{Field2}\includegraphics[scale=0.15]{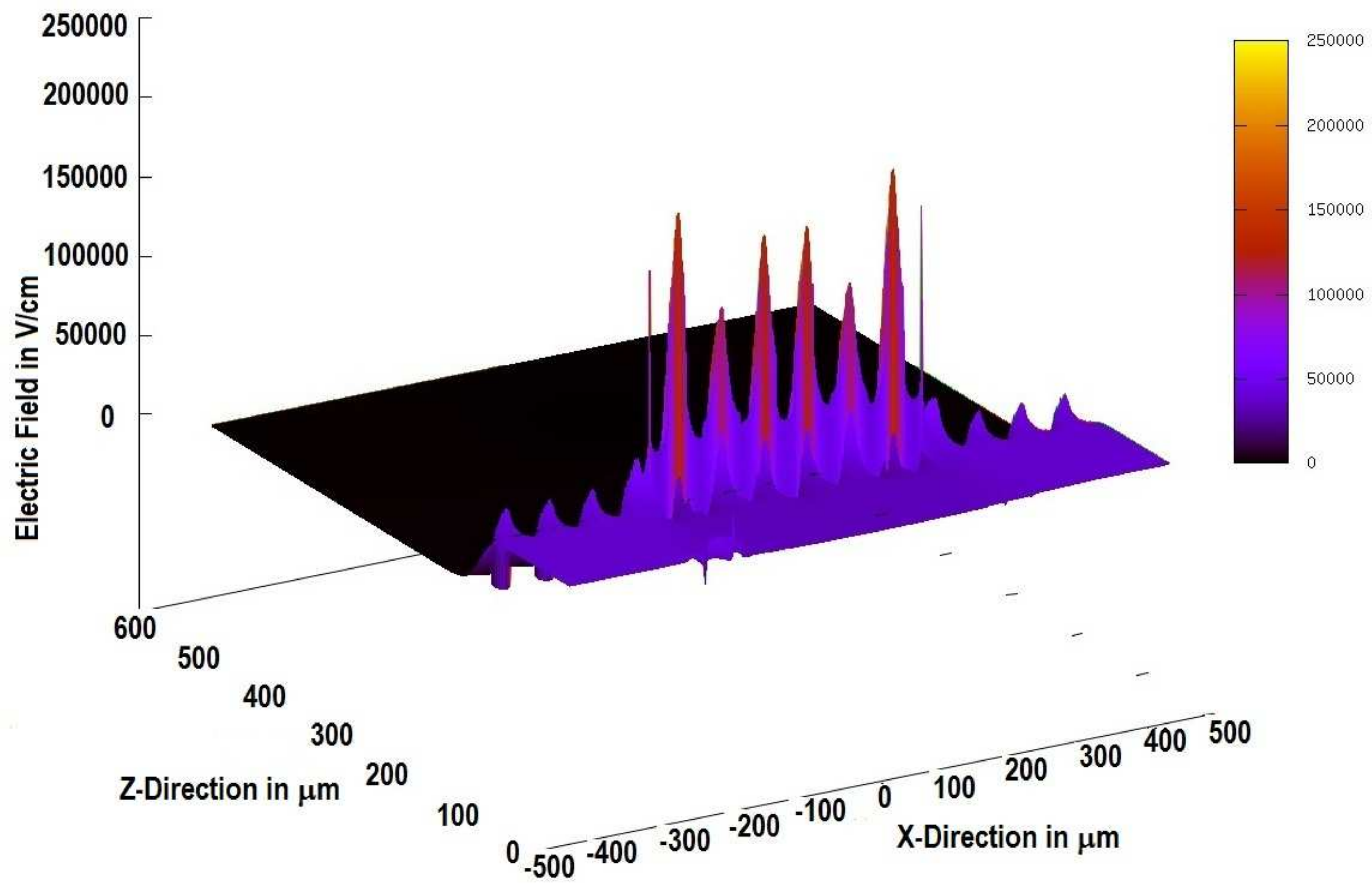}}
\subfigure[]
{\label{Field3}\includegraphics[scale=0.15]{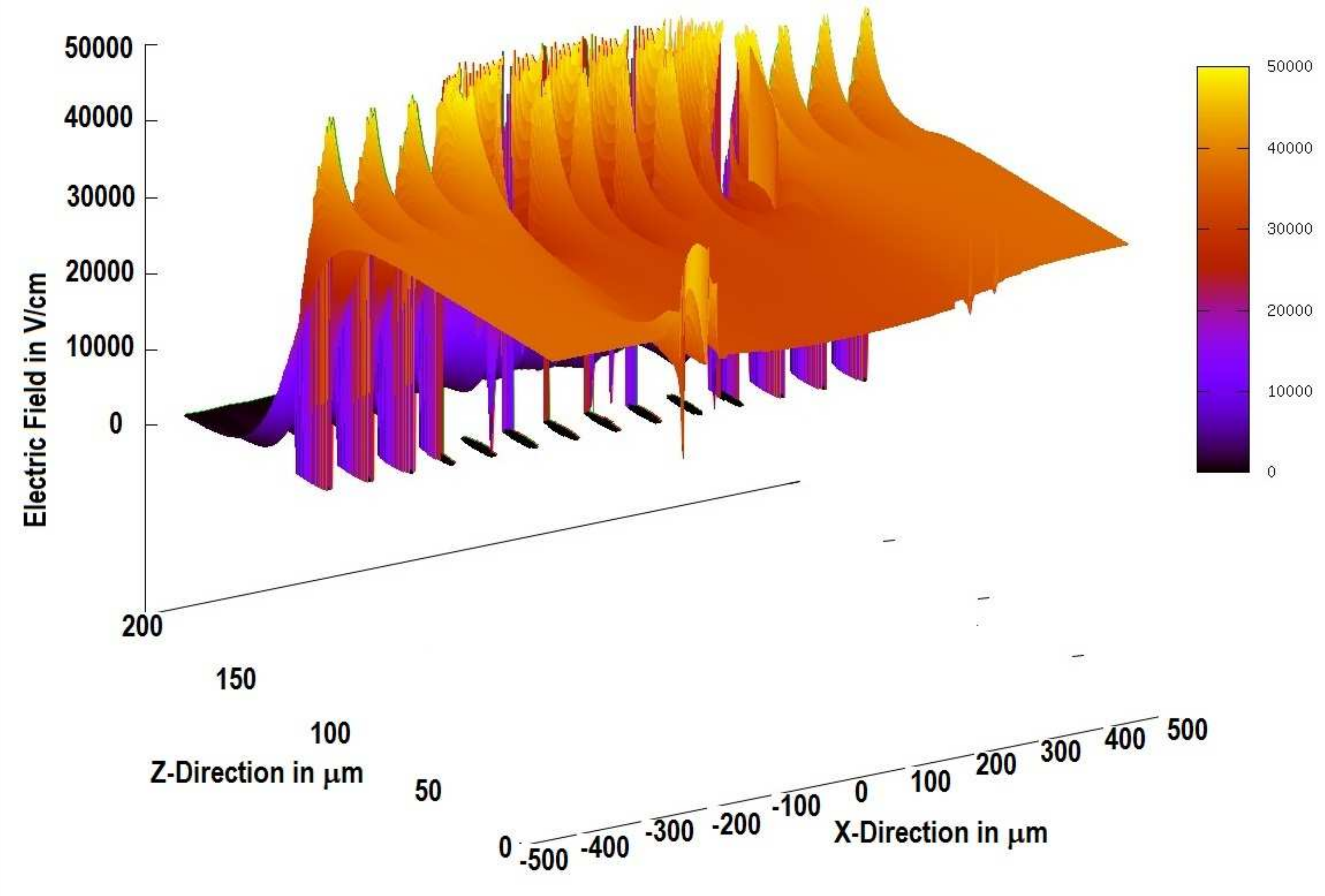}}
\subfigure[]
{\label{Field4}\includegraphics[scale=0.25]{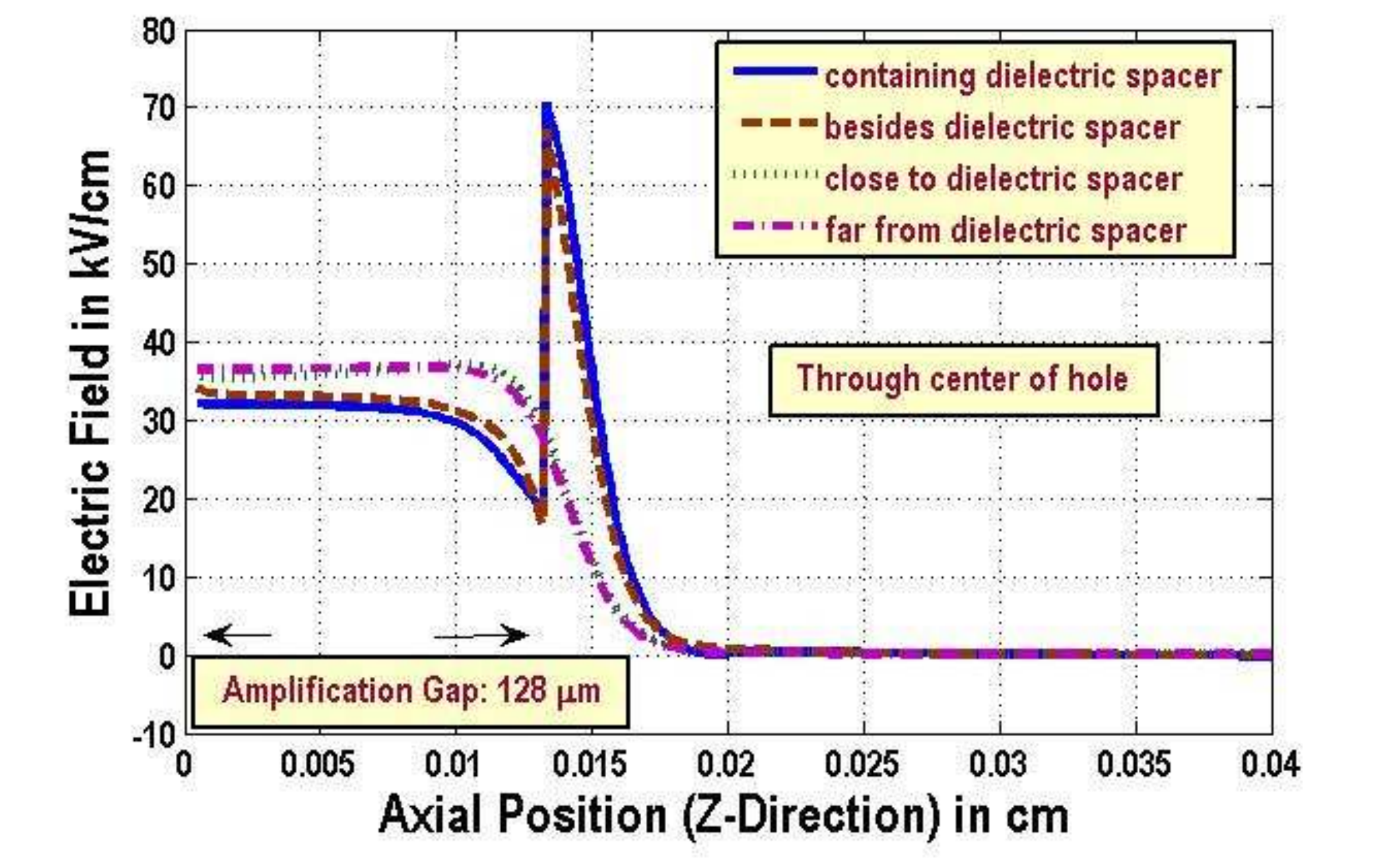}}
\caption{Electric field in XZ plane (a) without dielectric spacer, (b) with dielectric spacer, (c) close-up of (b); (d) along Z axis through different holes}
\label{Field}
\end{figure}

Introducing a full dielectric cylinder causes larger perturbation resulting in increased field values, particularly in the regions where the cylinder touches the mesh (Fig.~\ref{Field}).
The electric field through the mesh hole, near the spacer is also affected by the presence of this dielectric material.
As a result, the drift lines also get distorted.

\begin{figure}[ht]
\centering
{\includegraphics[scale=0.35]{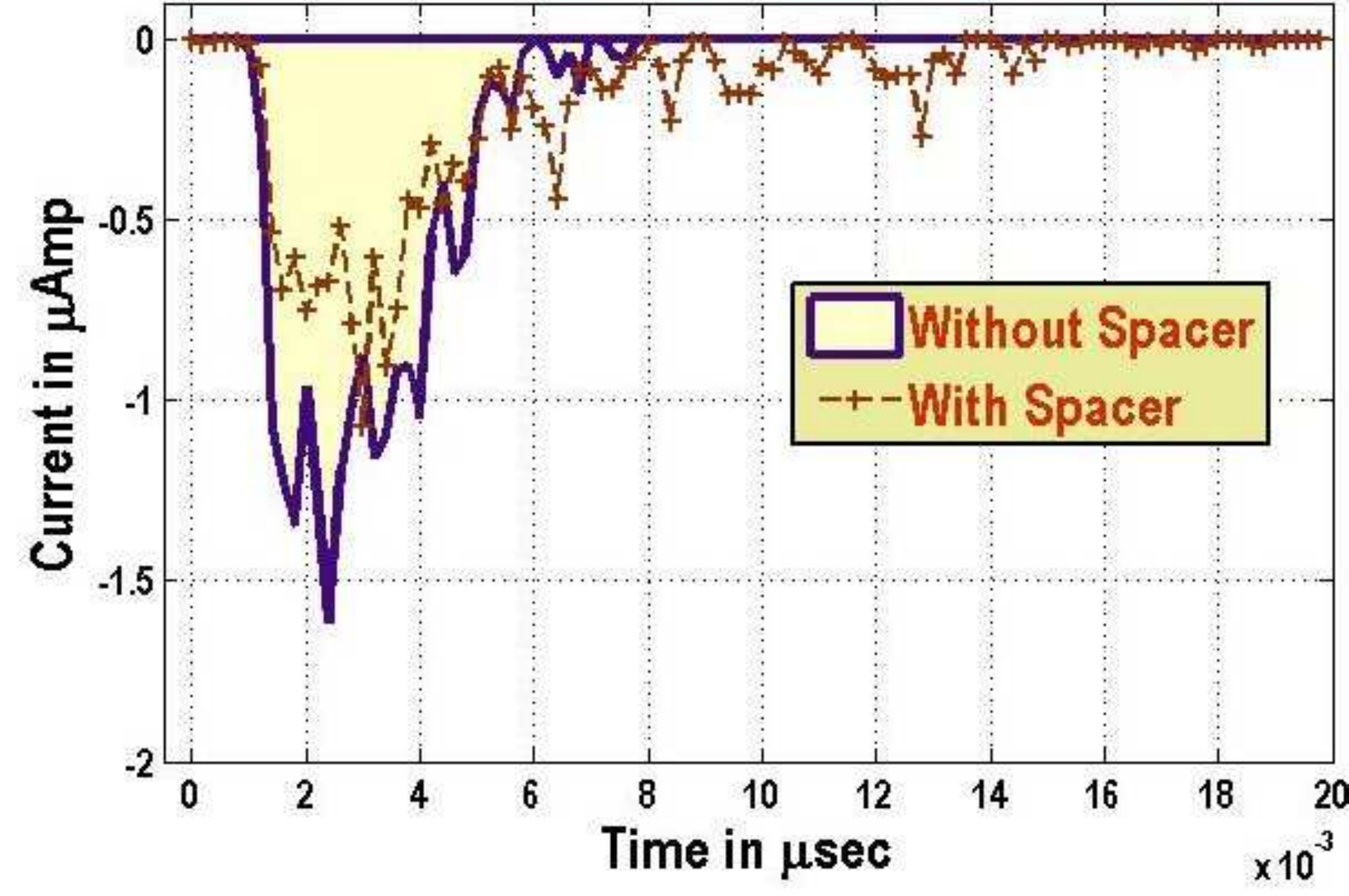}}
\caption{Effect of spacer on signal}
\label{Signal}
\end{figure}

A large number of electrons are lost on the spacer, resulting in reduced gain.
This effect is more prominent for the electrons originating from a track which is close to the micro-mesh (Table~\ref{Spacer}).
Due to this reduced gain, the signal strength also decreases whereas it has a longer tail resulting from distorted drift lines (Fig.\ref{Signal}).

\begin{table}[ht]
\caption{Electron Transparency and Gain (Without and With Spacer)}\label{Spacer}
\begin{center}
\begin{tabular}{|c||c|c|c||c|c|c|}
\hline
& \multicolumn{3}{|c||}{Without Spacer}& \multicolumn{3}{|c|}{WithSpacer}\\
\cline{2-4}
\cline{5-7}
Position of track above mesh&$25~\mu\mathrm{m}$&$50~\mu\mathrm{m}$&$100~\mu\mathrm{m}$&$25~\mu\mathrm{m}$&$50~\mu\mathrm{m}$&$100~\mu\mathrm{m}$\\
\hline
Electrons crossing mesh&97.794&97.304&97.549&97.549&95.343&95.833\\
\hline
Electrons reaching middle of amplification area&97.794&97.304&97.549&54.902&92.892&95.343\\
\hline
Gain&600&594&596&338&570&584\\
\hline
\end{tabular}
\end{center}
\end{table}

\section{Conclusions}

In this paper, we present a comparative study between bulk Micromegas having different amplification gaps.
Various detector characteristics such as gain, electron transparency, energy resolution have been measured experimentally.
Successful comparisons of these measured data with the simulation results indicate that the device physics is quite well understood.
The larger gap bulk Micromegas shows performance similar to a standard $128~\mu\mathrm{m}$ bulk.
Significant increase in the count rate has been observed for $192~\mu\mathrm{m}$ but this aspect needs further investigation.
Numerical estimation on the effects of spacers on detector characteristics indicated significant changes occurring around the spacer.

In future, further studies will be carried out using Micromegas having a wider range of amplification gaps in various conditions which will help users to choose an optimum amplification gap for a specific application.
We hope to make progress in other important areas such as calculation of ion back flow, Penning effect for a better understanding of these devices.

\section{Acknowledgment}

We thank our collaborators from ILC-TPC collaboration for their help and suggestions.
We also thank Rui de Oliveira and the CERN MPGD workshop for technical support.
This work has partly been performed in the framework of the RD51 Collaboration. 
We happily acknowledge the help and suggestions of the member of the RD51 Collaboration.
We are thankful to Abhik Jash for his help in some measurement and Pradipta Kumar Das and Amal Ghoshal for their technical help.
The work presented here has been partially financed by IFCPAR/CEFIPRA (Project No. 4304-1).
We thank our respective Institutions for providing us with the necessary facilities.

\end{document}